\documentclass[aps,prx,twocolumn,
               groupedaddress,superscriptaddress,
               10pt,amsfonts,amssymb,amsmath,
               citeautoscript,
               a4paper]{revtex4-2}

\usepackage{CJKutf8} 
\usepackage[UKenglish]{babel}
\usepackage{txfonts}
\usepackage{txfontsb}

\usepackage{graphicx}
\usepackage{placeins}
\newcommand{\yangchar}{\raisebox{-0.05em}{\includegraphics[height=0.85em]{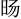}}}

\usepackage{mathrsfs}
\usepackage{physics}
\usepackage{stmaryrd}
\usepackage{gensymb}
\usepackage{xspace}

\usepackage{xr}
\usepackage{xcite}

\usepackage{xcolor}

\usepackage{siunitx}
\DeclareSIUnit[number-unit-product=]\percent{\char`\%}

\usepackage[pdftex]{hyperref}
\hypersetup{colorlinks,
            linkcolor={blue!75!black!80!yellow},
            citecolor={blue!75!black!80!yellow},
            urlcolor={blue!75!black!80!yellow},
            pdfstartview=FitH}

\makeatletter
\newcommand*{\addFileDependency}[1]{%
  \typeout{(#1)}%
  \@addtofilelist{#1}%
  \IfFileExists{#1}{}{\typeout{No file #1.}}%
}
\makeatother

\newcommand{\ui}{\mathrm{i}}
\newcommand{\ue}{\mathrm{e}}

\newcommand{\appropto}{\mathrel{\vcenter{
            \offinterlineskip\halign{\hfil$##$\cr
                \propto\cr\noalign{\kern2pt}\sim\cr\noalign{\kern-2pt}}}}}

\newcommand{\ie}{i.e.,\@\xspace}

\usepackage[nameinlink, capitalize]{cleveref}
\newcommand{\hnamecref}[1]{\hyperref[#1]{\namecref{#1}}}  
\crefname{figure}{Fig.}{Figs.}
\Crefname{section}{Sec.}{Secs.}
\crefname{equation}{Eq.}{Eqs.}
\Crefname{figure}{Figure}{Figures}
\Crefname{section}{Section}{Sections}
\Crefname{equation}{Equation}{Equations}
\newcommand{\hkuaffil}{\footnotesize Department of Physics and HK Institute of Quantum Science and Technology,\\The University of Hong Kong, Pokfulam, Hong Kong, China}
\newcommand{\natlabaffil}{\footnotesize State Key Laboratory of Optical Quantum Materials, \\The University of Hong Kong, Pokfulam, Hong Kong, China}

\begin{document}
\title{Observation of Non-Hermitian Skin Dynamics in the Liouvillian Regime}

\author{Shu~Yang~(\begin{CJK}{UTF8}{gbsn}杨树\end{CJK})}
\thanks{S.~Y. and Y.~S. contributed equally to this work.}
\affiliation{\hkuaffil}
\affiliation{\natlabaffil}
\author{Yeyang~Sun~(\begin{CJK}{UTF8}{gbsn}孙晔\end{CJK}\yangchar)}
\thanks{S.~Y. and Y.~S. contributed equally to this work}
\email{yysun47@hku.hk}
\affiliation{\hkuaffil}
\affiliation{\natlabaffil}
\author{Lingrui~Hong~(\begin{CJK}{UTF8}{gbsn}洪羚瑞\end{CJK})}
\affiliation{\hkuaffil}
\affiliation{\natlabaffil}
\author{Yi~Yang~(\begin{CJK}{UTF8}{gbsn}杨易\end{CJK})}
\email{yiyg@hku.hk}
\affiliation{\hkuaffil}
\affiliation{\natlabaffil}

\begin{abstract}
Open quantum systems generally do not perfectly preserve phase coherence: coupling to uncontrolled environments requires a density-matrix description based on the Liouvillian framework beyond pure-state wave evolution. 
Realizing and probing such dynamics in a programmable platform is therefore essential for connecting coherent physics to realistic dissipative settings. 
Here we implement a tunable open-system quantum walk in a photonic mesh lattice, where controlled phase noise produces adjustable dephasing and non-reciprocal gain-loss imbalance provides an independently tunable non-Hermitian drive. 
This allows us to continuously interpolate between coherent quantum walks and incoherent classical walks, and to observe how directional transport evolves in the Liouvillian regime. 
Using non-Hermitian skin dynamics as a probe, we measure the center-of-mass drift over both the coherence and non-Hermiticity parameters, revealing a crossover from coherence-enhanced to decoherence-enhanced transport in quantitative agreement with quantum-channel simulations.
We further program spatial and temporal interfaces to demonstrate interface accumulation and a long-time drift governed by the instantaneous channel. 
Our results establish a controllable photonic platform for simulating open quantum dynamics and show that decoherence can actively reshape non-Hermitian transport.
\end{abstract}

\maketitle

Coherence underlies a wide range of wave phenomena in ideal closed systems. 
In realistic scenarios, however, no system is perfectly isolated: coupling to surrounding environments is inevitable, leading to decoherence and dissipation~\cite{breuer2002theory, zurek2003decoherence}.
In quantum systems, decoherence arises from the leakage of phase information into the environment, thereby converting coherent superpositions into statistical mixtures~\cite{zurek1991decoherence, zurek2003decoherence}. 
Unlike closed systems, where wavefunctions evolve under Hermitian Hamiltonians, open quantum systems require a more universal density-matrix description governed by Liouvillian superoperators~\cite{lindblad1976generators, gorini1976completely,  breuer2002theory, diehl2011topology, verstraete2009quantum}. 
Because environmental coupling generally takes open quantum dynamics beyond Hermitian Hamiltonians, this Liouvillian formulation provides a natural bridge to the broader framework of non-Hermitian physics, which exhibits abundant phenomena absent in Hermitian systems~\cite{el2018non,ashida2020non,bergholtz2021exceptional,ding2022non}. 
Among these unique phenomena, non-Hermitian skin effect (NHSE) is the most typical and characteristic one, which reflects the anomalous sensitivity of spectra and eigenmodes to boundary conditions~\cite{yao2018edge,helbig2020generalized,weidemann2020topological,okuma2020topological,zhang2022review,okuma2023non,gohsrich2025non}. 
Under open boundary conditions (OBC), this sensitivity manifests as the macroscopic accumulation of an extensive number of bulk eigenmodes near the boundary, which reshapes the conventional understanding of bulk-boundary correspondence~\cite{yao2018edge,okuma2020topological,borgnia2020non,zhang2020correspondence,wang2024amoeba}. 
Discrete-time quantum walk is a versatile platform for studying non-Hermitian dynamics due to its controllable non-reciprocal transport.
Moreover, its discrete-time nature also makes it possible to monitor the stepwise redistribution of probability and the evolution of boundary-directed drift, as demonstrated extensively in photonic mesh lattices based on time multiplexing~\cite{regensburger2011photon,miri2012optical,wimmer2015observation,weidemann2020topological,weidemann2022topological,marques2023observation,yu2024dirac,feis2025space,pang2025topological, schreiber2010photons, schreiber2011decoherence, regensburger2012parity, boutari2016large, neves2018photonic}.
More recently, studies have begun to explore the interplay between the NHSE and environment-induced decoherence, especially in the fully incoherent limit where the skin dynamics is reduced to a biased classical Markov process~\cite{longhi2024incoherent}. 
Based on a polarization-encoded free-space photonic quantum-walk setup, NHSE-like directional transport was confirmed to persist and even be enhanced in the fully incoherent case compared with the coherent case~\cite{wang2026decoherence}. 
Whereas the previously investigated dynamics in both limits can be conveniently captured by state evolution operators, those in the intermediate regime of partial coherence, \ie the Liouvillian regime requiring a density-matrix superoperator description, remain experimentally absent.
Here, we report the direct observation of non-Hermitian skin dynamics in the Liouvillian regime by introducing phase noise controllability to the mesh-lattice platform.
The resulting stochastic phase fluctuations provide tunable dephasing, driving the system from coherent pure-state dynamics toward ensemble-averaged density-matrix dynamics governed by a Liouvillian superoperator.
We tune the dephasing strength by varying the noise amplitude and track the resulting skin dynamics through the long-time scaling of the center-of-mass (c.m.) displacement, from which the drift velocity is extracted.
We also exploit the programmability of the mesh lattice to introduce spatial and temporal interfaces, providing further probes of non-Hermitian skin dynamics under inhomogeneous channel evolution.
Beyond the coherent Hamiltonian and fully classical random walks, our results establish a direct experimental route for studying NHSE dynamics in the Liouvillian superoperator regime, and provide a versatile platform for exploring open quantum dynamics in controllable photonic systems.

\begin{figure*}[htbp]
    \centering
    \includegraphics[width=0.762\linewidth]{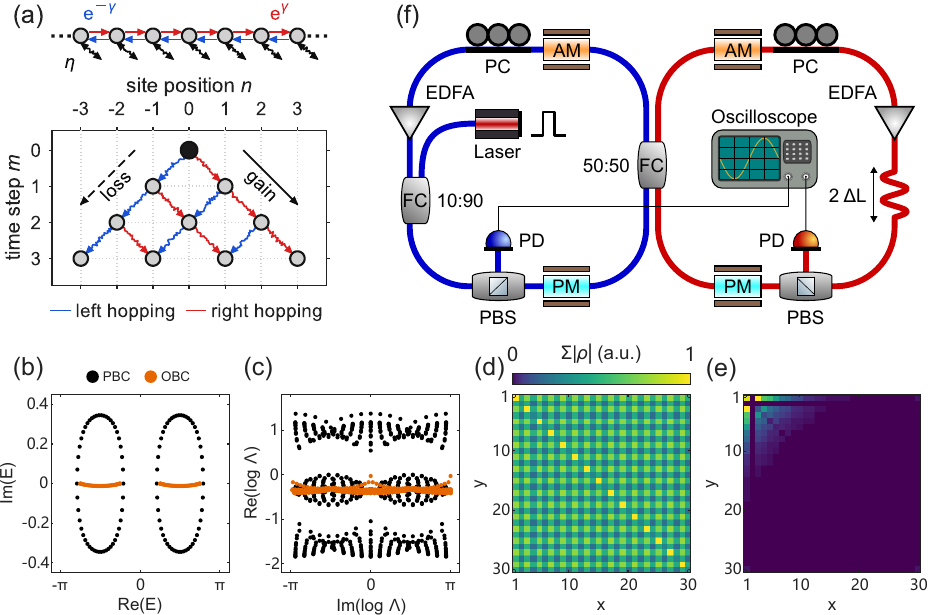}
    \caption{
            \textbf{Non-Hermitian quantum walk in the Liouvillian regime and its photonic implementation with tunable decoherence.} 
            (a) Schematic of the discrete one-dimensional non-reciprocal quantum walk model under decoherence. 
            The two directional components of the walker experience asymmetric amplification or attenuation controlled by the non-Hermiticity parameter $\gamma$. 
            Environmental coupling induces decoherence (indicated by trembling arrows) on the walker, which is characterized by the degree of coherence $\eta$.
            (b) Complex quasienergy spectra $E$ of the coherent step operator under PBC and OBC.
            The parameters are chosen as $\theta=\pi/4$ and $\gamma=1$. 
            (c) Corresponding spectra of the quantum channel, plotted in terms of $\log \Lambda$, where $\Lambda$ denotes the eigenvalue of the channel superoperator. 
            The semi-analytical simulations were performed for a lattice of size $2N=30$, with $\eta=0.72$ and the same $\theta$ and $\gamma$ as in (b).
            (d-e) Summed density distributions of the channel eigenmodes under (d) PBC and (e) OBC.
            (f) Experimental setup based on coupled fiber loops. 
            The laser wavelength is around \SI{1550.12}{\nano\meter}, and the pulse width is \SI{100}{\nano\second}. 
            FC, fiber coupler; EDFA, erbium-doped fiber amplifier; AM, amplitude modulator; PM, phase modulator; PD, photodetector; PC, polarization controller; PBS, polarization beam splitter. 
            The PBS-PM-FC-AM sections are polarization-maintained and slow-axis aligned, and the PMs are driven by uncorrelated noise signals.
    }
    \label{fig:fig1}
\end{figure*}
We start by considering a discrete-time quantum walk in a one-dimensional non-reciprocal lattice.
\Cref{fig:fig1}(a) illustrates the non-reciprocal lattice model and the quantum walk dynamics under single-site excitation. 
The lattice sites are labeled by $n=0,\pm1,\pm2,\ldots,\pm N$, and each site supports two internal states, denoted by $\ket{\rightarrow}$ and $\ket{\leftarrow}$, or equivalently, by a two-component spinor degree of freedom. 
The state vector of the system at the $m$-th time step can then be written as $\ket{\psi_m} = \sum_{n=-N}^{N} {\left( u_{n}^{m}\ket{n}\otimes\ket{\rightarrow} + v_{n}^{m}\ket{n}\otimes\ket{\leftarrow} \right)}$, where $u_{n}^{m}$ and $v_{n}^{m}$ are the probability amplitudes on the two internal states at site $n$.
In the absence of decoherence [indicated by the trembling arrows in \cref{fig:fig1}(a)], the dynamics of the quantum walk is governed by a step operator $\hat{U}=\hat{G}(\gamma)\hat{S}\hat{C}(\theta)$ with the internal states ordered as $\{\ket{\rightarrow},\ket{\leftarrow}\}$, 
which consists of a coin operator $\hat{C}=\sum_{n} \ket{n}\bra{n}\otimes\ue^{\ui\theta\sigma_x}$ that applies a local rotation to the internal states, 
a spatial shift operator $\hat{S}=\sum_{n} {\ket{n+1}\bra{n}\otimes\ket{\rightarrow}\bra{\rightarrow} + \ket{n-1}\bra{n}\otimes\ket{\leftarrow}\bra{\leftarrow}}$ that moves the walker to the right or left, 
and a non-Hermitian operator $\hat{G}(\gamma)=\sum_{n} \ket{n}\bra{n}\otimes\ue^{\gamma\sigma_z}$ which applies state-dependent amplification or attenuation to the internal states. 
Explicitly, the stroboscopic evolution of the walker is given by $\ket{\psi_{m+1}} = \hat{U}\ket{\psi_m}$, from which an effective non-Hermitian Hamiltonian can be defined through $\hat{U}=\ue^{-\ui\hat{H}_{\mathrm{eff}}}$~\cite{kitagawa2010topological, asboth2013bulk, rudner2013anomalous}.
The quasienergy spectrum is obtained from the eigenvalue equation $\hat{U}\ket{\psi_{\mu}}=\lambda_{\mu}\ket{\psi_{\mu}}$, with the effective complex quasienergy defined as $E_{\mu}=\ui\log{\lambda_{\mu}}$.
The NHSE is manifested in the sensitivity of the complex-energy spectrum to the boundary condition, especially through the contrast between periodic boundary condition (PBC) and open boundary condition spectra [\cref{fig:fig1}(b)]~\cite{yao2018edge,yokomizo2019non}.

\begin{figure*}[htbp]
    \centering
    \includegraphics[width=1\linewidth]{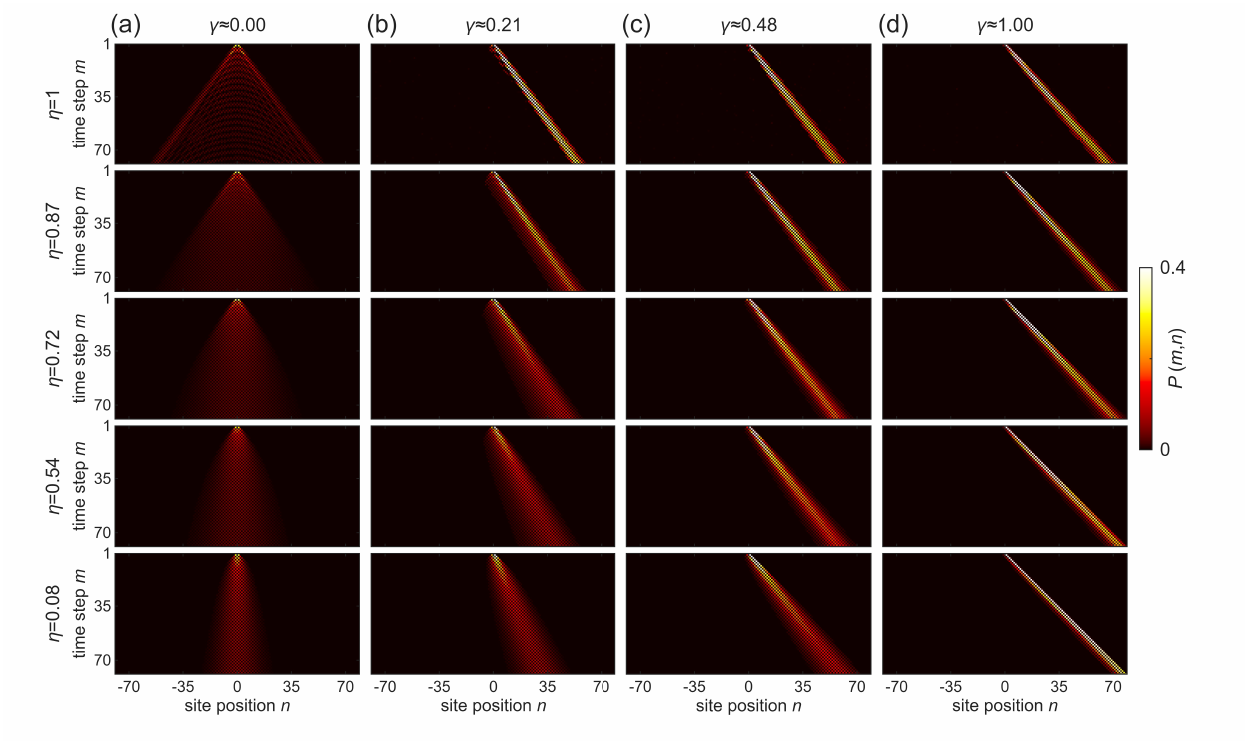}
    \caption{
        \textbf{Observation of non-Hermitian skin dynamics under tunable decoherence and non-Hermiticity.}
        (a-d) Measured normalized probability distributions $P(m,n)$ of the right-loop component for different combinations of $\eta$ and $\gamma$, where the degree of coherence $\eta$ is tuned by adjusting the amplitude of the Gaussian noise applied to the PMs, while the non-Hermiticity $\gamma$ is controlled by reducing the transmittance of the AM in the left loop.
    }
    \label{fig:fig2}
\end{figure*}
However, realistic quantum systems are often subject to environmental coupling that leads to decoherence. 
A paradigmatic example is pure dephasing, where the populations remain unchanged while the coherences between different basis states are gradually suppressed due to random phase fluctuations.
In the discrete-time quantum walk considered here, pure dephasing can be modeled by applying a stochastic phase operator $\hat{\Phi}_m = \sum_{n}{\ket{n} \bra{n} \otimes \mathrm{diag} ( \ue^{\ui\alpha_n^m},\ue^{\ui\beta_n^m})}$ at each time step, where $\alpha,\beta$ are random variables that follow a certain distribution $\mathcal{D}$ and fluctuate over site position $n$ and time step $m$.
Whereas a single realization of the walker under a known noise pattern can still be depicted by a state-vector representation $\ket{\psi_{m+1}} = \hat{U} \hat{\Phi}_m \ket{\psi_m}$, the random nature of the noise leads to a statistical mixture of different realizations, which cannot be captured by the pure-state description.
To account for this, we adopt a density-matrix formalism given by $\hat{\rho}_m=\overline{\ket{\psi_m}\bra{\psi_m}}$, where the statistical mixture generated by different noise realizations is written as an ensemble average over the pure states~\cite{breuer2002theory, nielsen2010quantum, song2019non}.
The evolution of $\hat{\rho}$ is described by a quantum channel $\hat{\rho}_{m+1} = \mathcal{E}(\hat{\rho}_m)$~\cite{kwiat2000experimental, brun2003quantum,brun2003quantumto, kovsik2006quantum,kendon2007decoherence, whitfield2010quantum,annabestani2010decoherence,broome2010discrete, schreiber2011decoherence,geraldi2019experimental,huang2026experimental}, which can be expressed in the Kraus representation as $\hat{\rho}_{m+1} = \eta\hat{U}\hat{\rho}_m\hat{U}^{\dagger} + (1-\eta) \sum_{s} \hat{U} \hat{K}_{s} \hat{\rho}_m \hat{K}_{s}^{\dagger} \hat{U}^{\dagger}$, where $\eta = \abs{\mathbb{E}_{\phi \sim \mathcal{D}}\!\left[\ue^{\ui\phi}\right]}^2$ and $0 \leq \eta \leq 1$ is the degree of coherence evaluated by the expectation $\mathbb{E}$ of the phase noise, and $\hat{K}_{s}=\sum_{n}{\ket{n}\bra{n}\otimes\ket{s}\bra{s}}$, $s\in\{\rightarrow,\leftarrow\}$ characterize the decoherence process~\cite{kraus1983states, nielsen2010quantum, broome2010discrete}.
Similar to the effective Hamiltonian defined from the coherent evolution operator, an effective Liouvillian superoperator $\mathcal{L}_{\mathrm{eff}}$ can be defined through $\mathcal{E}=\ue^{\mathcal{L}_{\mathrm{eff}}}$~\cite{lindblad1976generators, breuer2002theory}.
In this context, the relevant spectrum is that of the quantum channel, whose eigenmodes satisfy $\mathcal{E}(\hat{\rho}_{\nu}) = \Lambda_{\nu} \hat{\rho}_{\nu}$, and, after choosing a branch of the logarithm, the corresponding eigenvalues of the effective Liouvillian can be written as $\log\Lambda_{\nu}$.
This channel spectrum provides a natural framework for identifying dynamics in the Liouvillian regime.
\Cref{fig:fig1}(c) shows an example of the channel spectra under PBC and OBC, where the presence of NHSE is reflected in the distinct spectral shapes and the sensitivity to boundary conditions.
To further visualize this effect, we compute the eigenmodes of the quantum channel and sum the corresponding density matrices for the PBC and OBC cases, as shown in \cref{fig:fig1}(d) and (e), respectively.
Under PBC, the resulting density distribution remains extended over the lattice, whereas under OBC it is strongly localized near the boundary. 
This contrast provides direct evidence of skin localization in the Liouvillian regime.

\begin{figure*}[tp]
    \centering
    \includegraphics[width=1\linewidth]{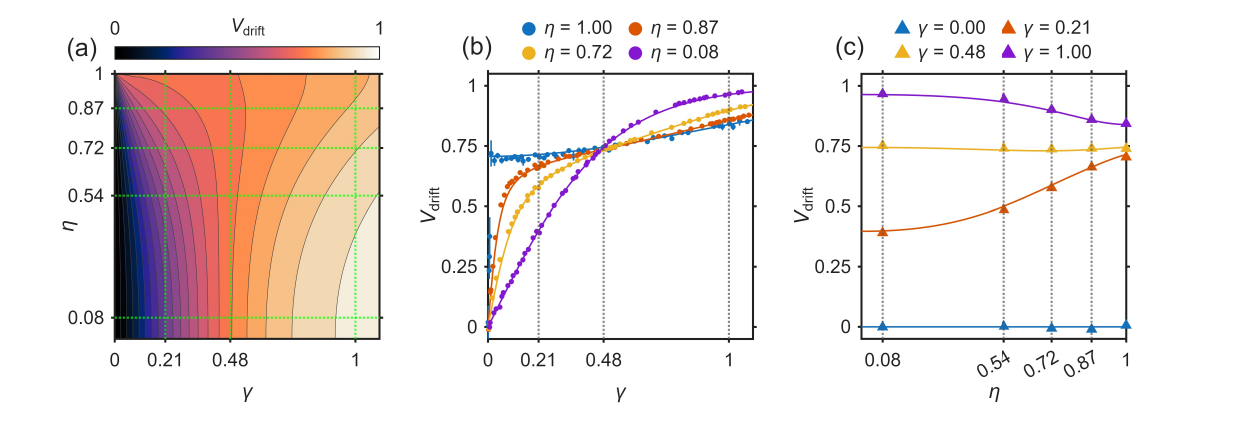}
    \caption{
        \textbf{Drift velocity of the non-Hermitian skin dynamics in the Liouvillian regime.}
        (a) Numerically simulated drift velocity $V_{\mathrm{drift}}$ (lattice constant and round-trip time are both chosen as unity) as a function of the non-Hermiticity parameter $\gamma$ and the degree of coherence $\eta$.
        The green dotted lines indicate the parameter cuts measured experimentally in (b) and (c).
        (b) Drift velocity as a function of $\gamma$ for several fixed $\eta$. 
        The solid curves are theoretical references.
        (c) Drift velocity as a function of $\eta$ for several fixed $\gamma$, corresponding to the experimental conditions used in \cref{fig:fig2}. 
        Most error bars are smaller than the marker size for all measurements.
    }
    \label{fig:fig3}
\end{figure*}
We have implemented the above theoretical model in a photonic mesh lattice setup shown in \cref{fig:fig1}(f), where the lattice sites are encoded by discrete time bins, and the two internal states, $\ket{\rightarrow}$ and $\ket{\leftarrow}$, correspond to the optical fields circulating in the right and left fiber loops~\cite{regensburger2012parity,weidemann2020topological}.
The two loops are connected by a fiber coupler (FC) that realizes the coin operator $\hat{C}(\theta)$, with the coin angle $\theta$ being controlled by the FC's splitting ratio. 
Throughout the experiments, we choose $\theta=\pi/4$ to maximize the walker spreading and the visibility of NHSE.
The right loop is $2\Delta L\approx \SI{64}{\meter}$ longer than the left loop, which gives a time delay of $\pm \Delta T\approx \SI{156.3}{\nano\second}$ for each loop with respect to their average round-trip time $\bar{T} \approx \SI{24.75}{\micro\second}$.
This realizes the spatial shift operator $\hat{S}$, where the walker moves to the right or left on the lattice depending on whether it is in the $\ket{\rightarrow}$ or $\ket{\leftarrow}$ state.
This delay also determines the effective lattice size $N=\lfloor\bar{T}/(2\Delta T)\rfloor=79$ that can be implemented in the experiment.
The non-Hermitian operator $\hat{G}(\gamma)$ is introduced jointly by an erbium-doped fiber amplifier (EDFA) and a DC-biased amplitude modulator (AM) in each loop. 
Since the amplitude transmittances of each loop $\kappa_{\mathrm{R, L}}$ cannot exceed unity (otherwise the system would be unstable), we define the non-Hermiticity parameter as $\gamma=\log\sqrt{\kappa_{\mathrm{R}}/\kappa_\mathrm{L}}$.
The decoherence is controlled by applying phase noise to the system, which is implemented by a pair of phase modulators (PMs) driven by different noise signals. 
Experimentally, we adopt Gaussian noise with a sampling rate of \SI{200}{\mega\hertz} (fast enough to ensure the noise is uncorrelated between different time bins), and the degree of coherence is characterized by its standard deviation $\sigma$ through $\eta=\ue^{-\sigma^2}$ (see Supplemental Materials). 
At this point, we have established a direct mapping between the theoretical model and the experimental setup, and the evolution equations of the system for a single realization are obtained as
\begin{subequations}\label{eq:evolution_equations}
\begin{align}
    u_{n}^{m+1} &= \ue^{+\gamma} \left ( u_{n-1}^{m} \cos{\theta} + \ui v_{n-1}^{m} \sin{\theta} \right ) \ue^{\ui \alpha_{n}^{m}}, 
    \label{eq:evolution_equation_u}
    \\
    v_{n}^{m+1} &= \ue^{-\gamma} \left ( v_{n+1}^{m} \cos{\theta} + \ui u_{n+1}^{m} \sin{\theta} \right ) \ue^{\ui \beta_{n}^{m}},
    \label{eq:evolution_equation_v}
\end{align}
\end{subequations}
where $\alpha,\beta \sim \mathcal{N}( 0,\sigma^2)$. 
Since the system is evaluated in a statistical manner, we define the (unnormalized) probability distribution of the walker as $\tilde{P}_{\rightarrow}{(m,n)} = \overline{\abs{u_{n}^{m}}^2},\tilde{P}_{\leftarrow}{(m,n)} = \overline{\abs{v_{n}^{m}}^2}$.
This is equivalent to the diagonal elements of the density matrix in the position and internal-state basis, \ie $\tilde{P}_s(m,n)=\mel{n,s}{\hat{\rho}_m}{n,s}$.

Experimentally, the system is initialized by injecting a single laser pulse into the left loop via a 10:90 FC, which corresponds to a single-site excitation $\ket{\psi_0}=\ket{n=0}\otimes\ket{\leftarrow}$ at the center of the lattice. 
The evolution of the walker is monitored by measuring the time-resolved pulse intensity in both loops, which gives direct access to the probability distribution of the walker's position and internal state at each time step.
After assigning each time-resolved pulse to a corresponding time bin, we denote the integrated intensity from the two PDs by $I_s(m,n)$ with $s\in\{\rightarrow,\leftarrow\}$ and define the normalized probability distribution as $P_s(m,n) = I_s(m,n) \slash \sum_{n^\prime}\left[I_{\rightarrow}(m,n^\prime)+I_{\leftarrow}(m,n^\prime)\right]$.
For this work, we focus on the signals measured from the right loop, which correspond to the projection on the $\ket{\rightarrow}$ state.
For notational simplicity, we write $P_{\rightarrow}(m,n)$ as $P(m,n)$ hereafter.

As a first experimental demonstration, we use the normalized distribution $P(m,n)$ to map out the quantum-walk dynamics under various combinations of the non-Hermiticity $\gamma$ and the degree of coherence $\eta$, as collected in \cref{fig:fig2}.
In the Hermitian limit $\gamma\approx 0$ [\cref{fig:fig2}(a)], the walker spreads almost symmetrically with respect to the initial site. 
The spreading profile changes from a ballistic interference pattern at $\eta=1$ to a smoother, diffusion-like distribution as $\eta$ decreases, showing the controlled suppression of quantum coherence by phase noise.
When positive $\gamma$ is introduced, the distribution develops a pronounced directional drift toward positive lattice sites. 
This drift becomes stronger as $\gamma$ increases [\cref{fig:fig2}(b)-(d)], indicating the accumulation of probability along the preferred hopping direction induced by the non-Hermitian gain-loss imbalance.
On the other hand, although decoherence washes out the interference fringes and broadens the wave packet, it does not eliminate the non-reciprocity-induced transport, as evidenced by the persistent drift even in the nearly incoherent regime $\eta \approx 0.08$.
More importantly, as $\gamma$ is further increased, the drift of decoherent walkers can be even more significant than that of the coherent ones, as shown in \cref{fig:fig2}(d). 
This trend connects naturally to previous theoretical~\cite{longhi2024incoherent} and experimental~\cite{wang2026decoherence} studies of the NHSE in the two limiting cases of fully coherent quantum walks ($\eta=1$) and fully incoherent classical walks ($\eta=0$). 
Here, the tunable phase noise further enables us to access the intermediate Liouvillian regime between these limits under arbitrary degrees of decoherence.

A standard way to quantify the directed transport is to track the evolution of the wave-packet center of mass (c.m.), defined as $\expval{n(m)} = \sum_{n} n P(m,n)$~\cite{rudner2009topological, schreiber2010photons, weidemann2020topological, longhi2024incoherent}.
The drift velocity is then extracted from the asymptotic slope of $\expval{n(m)}$ in the long-time limit,
\begin{equation} \label{eq:drift_velocity}
    V_{\mathrm{drift}} = \lim_{m \to\infty} \frac{ \delta \expval{n(m)}}{\delta m}.
\end{equation}
Specifically, in the two limiting cases of $\eta=1$ and $\eta=0$, explicit analytical expressions for $V_{\mathrm{drift}}$ can be obtained as $V_{\mathrm{drift}}^{\eta=1} = \operatorname{sgn}{\gamma} \cos{\theta} \cosh{\gamma}/\sqrt{1+\cos^2{\theta} \sinh^2{\gamma}}$ and $V_{\mathrm{drift}}^{\eta=0} = \cos^2{\theta} \sinh{(2\gamma)}/\sqrt{\sin^4{\theta} + \cos^4{\theta}\sinh^2{(2\gamma)}}$, respectively~\cite{longhi2024incoherent}.
In the intermediate Liouvillian regime ($0<\eta<1$), however, no closed-form expression for $V_{\mathrm{drift}}$ is available.
Nevertheless, semi-analytical results can be reached by iterating the quantum channel $\mathcal{E}$ to the initial density matrix $\hat{\rho}_0$, for a fairly large number of time steps until the asymptotic regime is reached, and then extracting $V_{\mathrm{drift}}$ from the c.m. evolution via \cref{eq:drift_velocity}.
\Cref{fig:fig3}(a) illustrates the simulated contour map of $V_{\mathrm{drift}}$ with respect to $\gamma$ and $\eta$, where the green dotted lines mark the experimental parameters used in \cref{fig:fig2,fig:fig3}(b)-(c).
The contour map reveals two distinct transport regimes: 
For weak non-Hermiticity $\gamma$, the coherent walker exhibits a larger $V_{\mathrm{drift}}$ than that of the decoherent one, so that $V_{\mathrm{drift}}$ increases with $\eta$; 
in contrast, for strong non-Hermiticity $\gamma$, the decoherent dynamics supports a larger drift velocity, and $V_{\mathrm{drift}}$ decreases as the system becomes more coherent. 
These two tendencies are separated by a crossover region around $\gamma \approx 0.48$, where $V_{\mathrm{drift}}$ becomes only weakly dependent on $\eta$.
Importantly, this transition from coherent to incoherent reveals a nontrivial Liouvillian response: decoherence can actively reshape and even enhance the non-reciprocal transport.
To compare the predictions with experiments, we extract $V_{\mathrm{drift}}$ from the measured c.m. trajectories using linear regression, and plot the results as cuts through the contour map [\cref{fig:fig3}(a)]. 
\Cref{fig:fig3}(b) shows $V_{\mathrm{drift}}$ as a function of $\gamma$ for several fixed values of $\eta$. 
Across all conditions of $\eta$, the measured drift velocity increases with non-Hermiticity $\gamma$ and follows the semi-analytical simulations, confirming that the directed transport remains well defined in the Liouvillian regime.
The fully coherent ($\eta=1$) data serve as a limiting reference and are consistent with the analytical analysis, while the partially coherent cases are also in good agreement with the semi-analytical simulations.
\Cref{fig:fig3}(c) summarizes the drift velocities extracted from the experimental data in \cref{fig:fig2}, and shows a complementary result of $V_{\mathrm{drift}}$ as a function of $\eta$ for several fixed values of $\gamma$. 
For $\gamma=0$, the drift vanishes for all $\eta$, as expected for a reciprocal lattice. 
For small positive $\gamma$, increasing coherence enhances the drift, whereas for large $\gamma$ the drift is stronger in the decoherent regime and gradually decreases with increasing $\eta$. 
Excellent agreement is reached between the experimental data and the theoretical results.

\begin{figure}[htbp]
    \centering
    \includegraphics[width=1\linewidth]{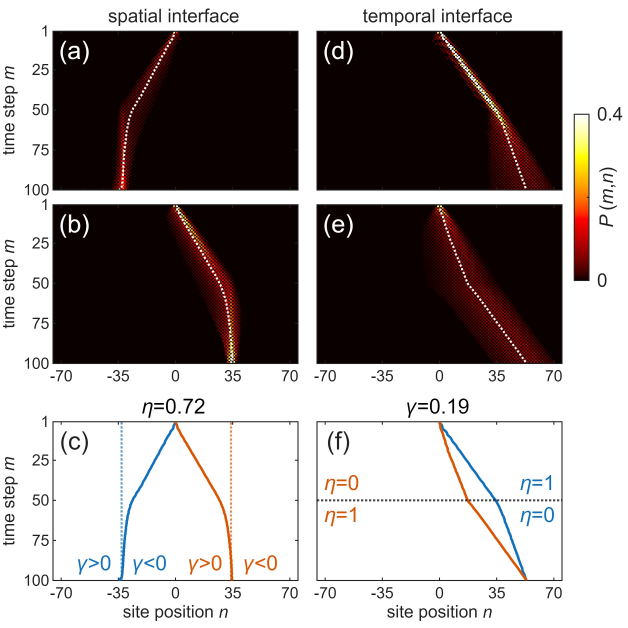}
    \caption{
        \textbf{Programmable spatial and temporal interfaces for the decoherent non-Hermitian skin effect.}
        (a-b) Measured probability distributions $P(m,n)$ under spatially dependent non-Hermiticity, where the sign of $\gamma$ is reversed across programmed interfaces at (a) $n=-35$ and (b) $n=+35$. 
        (c) Corresponding c.m. evolution for the two mirror-symmetric spatial-interface configurations, showing accumulation toward the programmed interfaces.
        (d-e) Measured probability distributions under temporal-interface protocols, where the degree of coherence is abruptly switched at $m=50$.
        (f) Corresponding c.m. evolution for the two temporal-interface configurations.
        The two trajectories reach nearly the same final displacement despite the reversed order of coherent and decoherent evolution.
    }
    \label{fig:fig4}
\end{figure}
Beyond measuring the drift velocity in homogeneous lattices, the programmability of our photonic mesh lattice setup further allows us to engineer interfaces in both space and time, and to investigate the corresponding Liouvillian dynamics.
We first create spatial interfaces by reversing the sign of the non-Hermiticity parameter $\gamma$ across a selected lattice site, while keeping the degree of coherence fixed at $\eta=0.72$.
As shown in \cref{fig:fig4}(a)-(b), two mirror-symmetric configurations are implemented, with the interfaces located at $n=-35$ and $n=+35$, where the walker is progressively attracted to the programmed interfaces, giving a real-space visualization of skin accumulation in the presence of decoherence.
This accumulation is also confirmed by the c.m. evolution shown in \cref{fig:fig4}(c), where the c.m. first drifts toward the interface and then remains localized near it within the experimental time window. 
The reversal of the accumulation direction between the two mirror-symmetric configurations further rules out a fixed directional bias of the setup and confirms that the localization is controlled by the programmed sign structure of $\gamma$.
Next, we exploit the temporal programmability of the platform to introduce a temporal interface, implemented as an abrupt switch of the degree of coherence at $m=50$ while keeping the non-Hermiticity fixed.
Similar to the spatial interface case, we implement two temporally mirror-symmetric configurations where the degree of coherence is switched from $\eta=1$ to $\eta=0$ or vice versa, as shown in \cref{fig:fig4}(d) and (e).
\Cref{fig:fig4}(f) shows the extracted c.m. evolution for both configurations, and remarkably, we find that exchanging the order of the coherent and decoherent evolution segments leads to nearly the same final displacement, even though the intermediate wavepacket profiles are different.
This order independence indicates that, for uncorrelated phase noise as in our case, the Liouvillian walker possesses an intrinsic long-time drift velocity determined by the instantaneous quantum channel, and is insensitive to the residual coherence inherited from earlier evolution.
Together, these interface experiments demonstrate the versatility of our platform for probing non-Hermitian skin dynamics in the Liouvillian regime, where non-Hermiticity and decoherence can be programmed independently in both space and time.

In summary, by introducing controlled phase noise into a non-reciprocal quantum walk, we realized a quantum-channel evolution in which the degree of coherence and the non-Hermiticity can be independently tuned. 
This allowed us to move beyond the two limiting cases of coherent quantum walks and incoherent classical walks, and to directly access the intermediate regime where decoherence reshapes the non-reciprocal transport. 
Through measurements of the wavepacket c.m. dynamics, we mapped out the drift velocity as a function of both the non-Hermiticity $\gamma$ and the degree of coherence $\eta$, revealing a crossover between coherence-enhanced and decoherence-enhanced transport. 
Furthermore, by exploiting the spatiotemporal programmability of our platform, we engineered spatial and temporal interfaces to probe the skin dynamics under spatial or temporal inhomogeneity. 
Our work establishes a flexible experimental platform for studying non-Hermitian transport under tunable decoherence and paves the way for exploring the Liouvillian skin effects~\cite{song2019non, longhi2020unraveling, liu2020helical, haga2021liouvillian, okuma2021quantum, yang2022liouvillian,kawabata2023entanglement,ekman2024liouvillian, hu2025many, longhi2026erratic} in broader dissipative settings, where environmental couplings can play an intrinsic role in reshaping the spectral localization.
\vspace{0.5cm}

\emph{Acknowledgments}---We thank Yu Xia and Chao Xiang for experimental support, and Jinlou Ma and Junlin Wang for helpful discussions.
\bibliographystyle{apsrev4-2}
\bibliography{reference}
\end{document}